\documentclass[conference]{IEEEtran}
\usepackage{mathtools, cuted} 
\usepackage{lettrine} 
\usepackage{amsthm,amsmath} 
\usepackage{epsfig,amssymb,amsbsy,verbatim,subfigure,array}
\usepackage{pstricks,cite,url}
\usepackage{multicol,afterpage,wrapfig,paralist} 
\usepackage{hyperref}
\usepackage{breakurl}
\usepackage{verbatim,tikz,subfigure}
\usepackage{tikz,pgfplots}
\usepackage[english]{babel}
 
{}

\def\rk#1{{\red\sffamily\small\em $\Rightarrow$  (RK) #1 $\Leftarrow$}}
\def\d{\mathrm{d}}
\renewcommand\qedsymbol{$\blacksquare$}

\title{An Energy Efficient Spectrum Sensing in Cognitive Radio Wireless Sensor Networks}
\author{\IEEEauthorblockN{Atchutananda Surampudi and Krishnamoorthy Kalimuthu} 
\IEEEauthorblockA{Department of Electronics and Communication Engineering,\\
SRM University,\\
Chennai, India 603203.\\
atchutasurampudi812@gmail.com, kalimuthu.k@ktr.srmuniv.ac.in}
}

\begin{document}
\maketitle

\begin{abstract} 
The cognitive radio wireless sensor networks have become an integral part of communicating spectrum information to the fusion centre, in a cooperative spectrum sensing environment. A group of battery operated sensors or nodes, sensing information about spectrum availability in the radio links, needs an energy efficient strategy to pass on the binary information. A proper routing method through intelligent decision making can be a good solution. In this paper, an energy efficient routing protocol has been introduced and a performance comparison is made with the existing system in terms of it's energy consumption. The proposed routing technique considers both uniform and non-uniform clustering strategies and proves to be an energy efficient method of communicaton. 
\end{abstract}

\section{Introduction} 
The cognitive radio wireless sensor network (CR-WSN) is a group of sensors, performing the task of sensing spectrum availability information in a cognitive radio (CR) environment \cite{mitola}. The spectrum sensing can happen in a cooperative or a non-cooperative way. In a cooperative spectrum sensing these sensor nodes communicate their respective information to an information sink called the fusion centre (FC), by forming an interconnected wireless network \cite{cr1}. At the FC the data is aggregated and further decision for utilization of spectrum is taken. So, once a wireless network is formed, the one bit information present with each node is routed along a wireless route with the help of routing protocols and associated algorithms defining the protocols. There have been several routing protocols proposed for routing information along these sensor networks \cite{crs1}, \cite{crs2}, \cite{crs3}. In this paper, the focus has been made on the routing strategy used by sensors which sense spectrum availability information in cognitive radio systems. \\ \par
The existing system of sending the binary sensing information is done by hierarchical clustering using the low energy adaptive clustering hierarchy (LEACH) protocol and there have been several developments on the same \cite{leach}, \cite{mecn}. In such systems, energy efficiency has been the most important key point of the network routing design. Here we introduce a novel routing protocol which combines with hierarchical clustering and inter cluster routing to communicate the information to the FC in an energy efficient manner. This research puts light on the energy demand for the sensors to send the sensed information to the FC directly or along a certain intelligently decided path. A greedy algorithm called the prim's algorithm \cite{prim} is used to form a low energy minimum spanning tree at every round of cluster formation. The rest depends on the elected cluster head to take an energy efficient decision to send the sensed information. In this analysis, both uniform and non uniform size of clustering have been considered. \\ \par

This paper describes the proposed methodology of the new routing protocol and compares the energy parameter with that of the existing protocol, the LEACH. This paper is organized as follows: In section II, the clustering strategy of existing system is described and the proposed method of communication protocol is analyzed in section III. Section IV discusses the results and comparisons made with the existing system. Finally the paper concludes with section V.

\section{Existing System}
A group of Wireless Sensors are placed for spectrum sensing along with the FC, where the data is aggregated. These nodes are further differentiated into advanced and normal nodes and are powered by their residual energy. A probability of selection of Cluster Head (CH) is pre-specified. The system undergoes a certain number of rounds and selection of CH is based on distance and energy parameters of each node from the FC. From \cite{ch}, the condition for electing the CH for each round is given as

\begin{equation}
T(s) = \left\{
               \begin{array}{l}
                 \frac{p}{1-p(r mod \frac{1}{p})}; s \in G, \\ \\
                 0,  otherwise, 
              \end{array}
              \right. 
\label{eqn:rect5}   
\end{equation}
where $r$ is the current round of clustering, G is the set of nodes that have not been CHs in the last 1/p rounds. Each sensor node decides independently of other senor nodes whether it will claim to be a CH or not, by picking a random value between 0 and 1 and comparing it with a threshold T(s) based on a user-specified probability p. After the CH is selected, it aggregates information from its cluster of nodes and calculates its distance from the fusion centre. This distance is used to calculate the cost to send the one bit information to the fusion centre soon after election and the sensing information is sent. In this method, all elected CHs transmit the aggregated data to the FC directly. If the CH is far away to the FC, then it needs more energy to transmit. So, the overall lifetime becomes very short and a considerable number of nodes die after certain rounds.

\section{Proposed Method}
In the proposed method an energy efficient routing protocol has been introduced. As a further extension of the previously described method, the proposed method introduces the process of intelligent decision making and forwarding the information along a minimum energy route by forming a minimum spanning tree (MST) using a greedy algorithm called the Prim's algorithm. A greedy algorithm formulates a certain path in a stepwise manner by greedily choosing the next step with respect to the previous one, according to the least value of the available metric. The Prim's algorithm finds an MST in a connected, weighted graph.
It begins with a vertex and no edges and then applies the greedy rule, \\ \par
\textit{Add an edge of minimum weight that has one vertex in the current tree and the other not in the current tree}. \\ \par 
Using this Prims algorithm, the proposed communication is made as follows
\begin{itemize}
\item Node is elected as a CH using \eqref{eqn:rect5}. But it doesn't directly send the sensed information to the FC soon after election.
\item At once all the required 'M' number of CH's are elected for the 'rth' round and each CH now calculates it's Cartesian distance from the fusion centre and distances from every other node.
\item Using this information an adjacency matrix A(i,j) is formed, with distance as the metric.
\item MST is formed using the Prims algorithm. This connects all the elected 'M' number of CH's along a minimum energy route in that round. This tree varies dynamically with every round.
\end{itemize}
Now, as a development with respect to the earlier protocol, each CH maintains a sensing table which provides a comprehensive information about the one bit information sensed by itself and all other CH's that are present. So, the CH sends 'M' bit information rather than one bit along an intelligently decided path. It calculates the respective cost $(C)$ to the fusion centre and to the nearest CH along the minimum spanning tree, using the previously calculated distances as 
\begin{equation}
C = \left\{
               \begin{array}{l}
              (E_{t}+E_{d})M + E_{f}Md^{2}; d\leq d_{o}, \\ \\
              (E_{t}+E_{d})M + E_{m}Md^{4}; d> d_{o}.
              \end{array}
              \right. 
\label{eqn:rect51}   
\end{equation}
where $E_{t}$ is energy spent by transmitting one bit information, $E_{d}$ is data aggregation energy, $d_{o}$ is relative distance, $E_{f}$ and $E_{m}$ is energy spent for free space path loss and multipath loss respectively. Now the CH takes a decision on which route to select, either along the minimum spanning tree or directly to the fusion centre, based on the costs calculated. The minimum cost path is selected. So, all the CHs perform intelligent decision making and finally the fusion centre acquires complete information about the available spectrum in a short period of time. Moreover network life time is extended and comparatively very few nodes die after the same number of rounds considered earlier. The energy expended during transmission $E_{t}$ and reception $E_{r}$ for $M$ bits to a distance $d$ between transmitter and receiver for the secondary user is given by
\[ E_{t}=E_{e}M+E_{a}\alpha d. \]
\[ E_{r} = E_{e}M. \]
where $\alpha$ is path loss component, $E_{a}$ is energy constant for propagation, $E_{e}$ is the electronics energy. The following assumptions are made while making the analysis,
\begin{itemize}
\item A secondary users' network is stable and consists of one FC, one primary transmitter and $N$ number of cognitive radio users or sensor nodes.
\item The base station acts as the Fusion Centre and it has the location information of all secondary users, possibly determined using Global Positioning System (GPS).
\item The instantaneous channel state information of the channel is available at the each secondary user.
\item The channel between any two secondary users in the same cluster is perfect since they are close to each other.
\item In this research work, minimum tree cost determination is made with the adjacency matrix of distances.
\end{itemize}

\section{Simulation Results}
The performance of the proposed algorithm is carried with the following parameters given in Table I. The performance of the proposed routing protocol is compared with the conventional protocol. \\ \par
\begin{table}
\caption{PARAMETERS : This table shows the Parameters considered in this study.}
\label{abc}
%
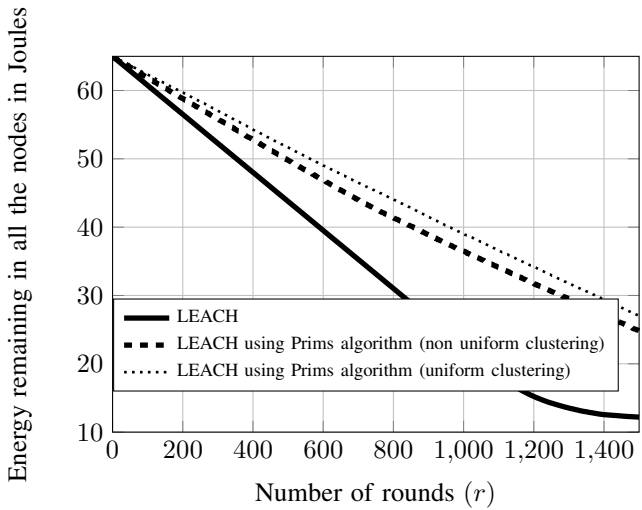
\caption{This graph shows the total energy remaining in Joules, in all the alive nodes $N$, at the end of each round $r$. Uniform and non-uniform clustering strategies have also been compared for the proposed protocol.}
\label{sim2}
\end{figure} 

\begin{figure}[ht]
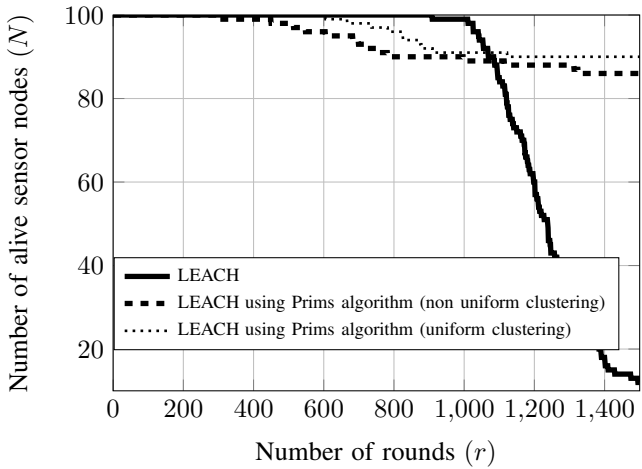

\centering
%
%
\caption{This graph shows the number of alive sensor nodes $N$, at the end of each round $r$. Uniform and non-uniform clustering strategies have also been compared for the proposed protocol.}
\label{sim2}
\end{figure} 

In the uniform clustering method, the number of clusters is restricted to 10. From Fig. \ref{sim1}, the proposed communication protocol proves to be energy efficient by $1.4$ times $($for $1500$ rounds$)$ and this increases the lifetime of the network. Moreover, in this research, when uniform clustering is restricted to $10$ CHs, the network residual energy is still higher by $27\%$ $($for $1500$ rounds$)$ than non-uniform clustering $($where the cluster size is not uniform for every cluster$)$. Thus, it is observed that the overall network residual energy using proposed protocol with uniform size clustering is much greater than the existing system and non-uniform clustering as well. From Fig. \ref{sim2}, it is observed that the number of active nodes after $1500$ rounds of clustering and spectrum sensing, in the proposed protocol with uniform clustering is more than the number of active nodes in the existing protocol by 7 times. So, with this it can be shown that the Proposed method is an energy efficient strategy with respect to the existing system of communication for CR-WSN. 
 
\section{Conclusion}
The analysis and simulation results show that the proposed protocol with uniform or non-uniform clustering shows better performance in terms of energy consumption. Moreover, this scales down a large size of network and proves to be an elegant design paradigm of reducing energy consumption. Moreover, all the elected CHs, along with the FC have complete information about the spectrum availability and as a result, network convergence has taken place. When $M$ number of bits are sent only by those CHs who are nearer to FC, the problem of channel congestion is also eliminated. As an extension work, the sensing tables can be used to store control information as well as to information storage and intelligence for sending the sensed information. This may be useful in case of mission critical situations like natural disasters, where network failure is a common phenomenon.


\end{document}